# An Ensemble Approach for Research Article Identification: a Case Study in Artificial Intelligence

Min Lu

*Hangzhou Science and Technology Information Institute,*

*Hangzhou, Zhejiang, China*

*lumin@hznet.com.cn*

Lie Tang

*Hangzhou Science and Technology Information Institute,*

*Hangzhou, Zhejiang, China*

*tanglie23@163.com*

Xianke Zhou

*Institute of Computer Innovation, Zhejiang university,*

*Hangzhou, Zhejiang, China*

*xkzhou@zjuici.com*

This study presents an ensemble approach that addresses the challenges of identification and analysis of research articles in rapidly evolving fields, using the field of Artificial Intelligence (AI) as a case study. Our approach included using decision tree, sciBERT and regular expression matching on different fields of the articles, and a SVM to merge the results from different models. We evaluated the effectiveness of our method on a manually labeled dataset, finding that our combined approach captured around 97% of AI-related articles in the web of science (WoS) corpus with a precision of 0.92. This presents a 0.15 increase in F1 score compared with existing search term based approach. Following this, we analyzed the publication volume trends and common research themes.We found that compared with existing methods, our ensemble approach revealed an increased degree of interdisciplinarity, and was able to identify more articles in certain subfields like feature extraction and optimization. This study demonstrates the potential of our approach as a tool for the accurate identification of scholarly articles, which is also capable of providing insights into the volume and content of a research area.

# Introduction

The total output of scientific research has seen a significant increase in recent years, with global scientific publishing output being 53% higher in 2020 compared to 2010. This growth is even more pronounced in emerging research fields, particularly in areas like artificial intelligence, robotics energy and materials science[1]. Emerging research fields, characterized by complex boundaries and rapid evolution[2], present unique challenges for accurate identification of research articles. Despite these difficulties, research on emerging fields holds great importance in understanding the evolution and impact of new technologies and scientific domains. For example, understanding the field of AI is crucial due to its interdisciplinary nature and its significant impact on various sectors ranging from economics to social development. Artificial Intelligence (AI), integrating elements from computer science, biology, psychology, and more, plays a pivotal role in advancing technologies like speech recognition, image processing, and intelligent robotics, thereby revolutionizing labor efficiency and creating new job demands[3]. By examining the various aspects of these fields, researchers and policymakers can gain valuable insights into their growth, trends, and potential implications for society.

The development of AI began in the 1950s, marking significant milestones such as the creation of neural networks and the introduction of terms like 'artificial intelligence' and 'machine learning.' More recent developments in the 2000s include IBM Watson, facial recognition technology, and autonomous vehicles[4]. Artificial Intelligence (AI) has gained significant attention in recent years, becoming a popular topic across various research fields. AI encompasses a wide range of techniques and methods that allow machines to perform tasks that typically require human intelligence. These tasks include, but are not limited to, problem-solving, learning, perception, and language understanding[5]. The advancements in AI have led to a plethora of sub-fields, such as machine learning, natural language processing, computer vision, and robotics[2].

AI's impact is widespread across industries. In healthcare, AI contributes to risk management, analytics, and knowledge creation, aiding in complex diagnostics and disease prevention[6]. In the public sector, AI is used to make cities safer and cleaner, aiding in traffic, pollution, and crime prediction[7]. Retail has seen AI applications in predictive analytics and customer service, enhancing operational efficiency and customer experience[8]. AI's contribution to manufacturing includes research and development, predictive analytics, and real-time operations management, notably impacting industrial automation[9]. These applications highlight AI's transformative potential in augmenting human abilities and driving significant economic growth. As a result of the growing interest in AI, the volume of AI-related articles has grown exponentially[10], highlighting the importance of tools for the accurate identification of AI-related articles. However, accurately identification within the vast corpus of scientific literature is challenging due to AI's broad, fuzzy, and rapidly-changing nature[2].

In the realm of research article classification, conventional keyword-based searches are a standard approach. For example, in the field of dentistry, a comprehensive review highlights the importance of correctly using keywords and Boolean operators for effective literature searches[11]. Similarly, in a study on acute respiratory tract infections (ARTI) and human metapneumovirus (hMPV) in children, researchers used a title and abstract (TIAB) search strategy with major keywords on hMPV infections to extract relevant data from various international publications. This approach helped in analyzing the epidemiological aspects of hMPV infections in children globally[12]. Both examples underscore the significance of keyword-based searches in efficiently sifting through and classifying large volumes of academic literature across diverse fields.

Another conventional method for classifying research articles involves using generic classification systems, such as the Web of Science (WoS) categories and Scopus categories. For instance, a bibliometric analysis in orthopedics utilized the WoS category to identify influential studies in the field, demonstrating the application of these systems for academic analysis[13]. Similarly, a study reviewing journal categories in WoS, Scopus, and MathSciNet bases under the title quartiles illustrated the use of these databases for journal ranking and classification, highlighting the significance of such systems in evaluating and categorizing scientific literature[14]. These examples show the prevalence of generic classification systems in scholarly research, offering a structured approach to categorize and evaluate academic publications.

Deep learning techniques have also been increasingly employed in the classification of scholarly articles. A study using convolutional neural networks (CNN) for scientometric analysis and classification demonstrated this approach's effectiveness. By incorporating publication, author, and content features, both explicit and implicit, into CNN models, this method achieved higher precision, recognition, and F1-score than traditional machine learning methods[15]. Another research utilized transformer models like BERT, specifically SciBERT, for text classification of peer review articles. This study found that sentence embeddings obtained from SciBERT, combined with entity embeddings, significantly enhanced classification performance, demonstrating the potency of deep learning methods in this domain[16].

Our approach of identifying AI-related research articles adopts an ensemble model, which synergizes the strengths of various techniques. We employ a decision tree algorithm to categorize articles based on a hierarchical set of Web of science categories. Alongside this, we utilize SciBERT, a BERT language model specifically trained on scientific literature, to analyze and understand the context and content of the articles. In addition to these methods, our model incorporates keyword matching across different fields. By scanning articles for a comprehensive set of AI-related keywords, we ensure that relevant publications are not overlooked, especially those that might be on the periphery of traditional AI categories. Finally, we used a Support Vector Machine (SVM) to merge the results from the decision tree, SciBERT, and keyword matching and providing a final classification. This combined approach enhances our model's ability to accurately identify AI-related articles amidst the vast and rapidly evolving corpus of scientific literature.

We manually labeled set of 4000 articles as either AI-related or non-AI for training, and these labels to calculate the precision and recall of our approach by 5-told cross validation. The results yielded a precision of 92% and recall of 97%.

# Literature review

### Research article classification in emerging fields

Bibliometric methods are frequently employed in the classification of research articles, particularly in emerging fields. These methods involve a comprehensive analysis of publication patterns and trends.For instance, in synthetic biology, a study outlined a bibliometric approach to define the field by analyzing a core set of papers and refining keywords identified from these sources, also including articles from dedicated journals[17].

In nano-energy research, Guan et al.[18] used bibliometric and social network analysis to investigate the exponential growth of research output and compare scientific performances across countries. This study revealed a shift in the global share of nano-energy research, with emerging economies like China showing significant development momentum. The research also highlighted the evolving patterns of scientific collaboration in this field, marking a shift in influence from traditional scientific powerhouses to emerging economies.

The detection of emerging research topics (ERTs) using bibliometric indicators was the focus of another study. This research distinguished ERTs from common-related topics and developed a method to uncover high-impact ERTs with a fine level of granularity. The study not only identified ERTs but also proposed different research and development strategies for each topic, emphasizing the future economic and social impact of these ERTs and the reduction of uncertainty in research directions[19].

### Search strategies of Artificial intelligence

Previous efforts to classify articles related to Artificial Intelligence (AI) have employed a variety of strategies. A simple yet commonly used method involves search strategies such as TS=("artificial intelligence"), which is straightforward but may miss nuanced aspects of AI research. For instance, a study examining AI developments in China used this strategy to analyze the interaction between academic research and policy-making, providing insights into the evolving relationship between these domains[20].

More complex approaches involve refining and complementing keywords with subject categories. Another study adopted a bibliometric definition for AI, starting with core keywords and specialized journal searches. This method was then enhanced by extracting high-frequency keywords from benchmark records, which was compared with other search strategies to profile AI's growth and diffusion in scientific research. This approach allowed for a more detailed understanding of AI's multidisciplinary development and the contributions from diverse disciplines[21].

In addition, the World Intellectual Property Organization (WIPO) in 2019 applied a comprehensive search strategy that combined patent classification codes with an extended list of keywords. This strategy was based on a thorough literature review, established hierarchies, web resources, and manual checking. To identify AI-related publications, approximately 60 words or phrases specific to AI concepts were queried across all subject areas in the Scopus scientific publication database. Additionally, about 35 words or phrases related to AI were applied specifically to the Scopus subject areas of Mathematics, Computer Science, and Engineering[2].

**Deep learning models in research article classification**

Deep learning models have also been utilized in identifying AI-related articles and patents. Dunham et al. [22] described a strategy leveraging the arXiv corpus to define AI relevance, employing deep learning techniques like convolutional neural networks. This approach yielded high classification scores and precision in identifying AI-relevant subjects, demonstrating the effectiveness of supervised solutions in updating AI-related publication identification in line with research advancements. Miric et al.[23] demonstrated the use of machine learning tools, including deep learning, for classifying unstructured text data. The study focuses on identifying AI technologies in patents, comparing ML methods with traditional keyword-based approaches. The findings indicate the superiority of ML in terms of accuracy and efficiency, underscoring the advantages of these methods in capturing the complex and evolving nature of AI innovation. Siebert et al.[24] utilized both supervised and unsupervised machine learning approaches for analyzing a large corpus of AI-related scientific articles. This method involved keyword extraction and seeding, followed by machine learning-based optimization and clustering to identify and analyze trends within AI research.

Search strategy-based classification methods, while convenient, often have limitations in terms of recall, potentially missing relevant articles that do not fit neatly into predefined keyword categories. Traditional machine learning methods for classifying AI-related articles, such as using simple models like Random Forest or SciBERT, also face performance constraints. They may not fully capture the complexity and nuances of AI research due to their reliance on existing models.

In contrast, our approach integrates keyword matching, decision tree based category classification, and the advanced capabilities of SciBERT. This combination allows for a more nuanced and comprehensive classification of AI-related articles, enhancing both accuracy and efficiency. By leveraging the strengths of each method, this integrated approach aims to provide a more robust and effective solution for classifying AI-related scholarly works.

**Decision Tree**

Decision trees are a fundamental machine learning technique used for classification and regression tasks. They operate by breaking down a dataset into smaller subsets while simultaneously developing an associated decision tree incrementally. The final result is a tree with decision nodes and leaf nodes, where each internal node represents a "test" on an attribute, each branch represents the outcome of the test, and each leaf node represents a class label[25].

The primary formula for decision tree algorithms involves the concept of information gain, which is derived from entropy. Entropy, a measure of disorder or uncertainty, is calculated using the formula:

$$\text{Entropy(S)} = -\sum_{\text{i=1}}^{\text{n}} p_i \log_2 p_i$$

Where $p_i$ is the proportion of the number of elements in class i to the number of elements in set S. Information gain, then, is defined as the difference in entropy before and after a dataset is split on an attribute. It is given by:

$$\text{Information Gain(S, A)} = \text{Entropy(S)} - \sum_{v \in \text{Values(A)}} \frac{|\text{S}_v|}{|\text{S}|} Entropy(S_v)$$

where A is the attribute and $S_v$ is the subset of S for which attribute A has value $v$.

**SciBERT**

BERT (Bidirectional Encoder Representations from Transformers) works on the principle of the Transformer, an attention mechanism that learns contextual relations between words in a text[26]. Unlike traditional models that read text sequentially (left-to-right or right-to-left), BERT reads the entire sequence of words at once. BERT is pre-trained on a large corpus of text and then fine-tuned for specific tasks. The Transformer model uses an attention mechanism that weighs the influence of different words on each other:

$$Attention(Q, K, V) = \text{softmax}(\frac{QK^T}{\sqrt{d_k}})V$$

Here, $Q$, $K$, $V$ are queries, keys, and values respectively, and $d_k$ is the dimension of the keys.

BERT uses multiple layers of the Transformer, and each layer outputs transformed representations of the input text.

SciBERT adapts the BERT model specifically for scientific texts. It is pre-trained on a large corpus of scientific literature, encompassing a wide range of domains[27]. This specialization allows SciBERT to better understand and process the language used in academic and technical documents. It has been proven effective in tasks like classification[16], entity recognition[28], and relationship extraction[29] in scientific texts. As is mentioned before, it has been utilized in the classification of AI-related scientific literatures and patents[22, 23]. SciBERT's ability to grasp complex scientific concepts makes it particularly useful in classifying and analyzing scholarly articles, including those related to AI research.

**SVM**

Support Vector Machine (SVM) is a supervised machine learning algorithm used for both classification and regression. SVM works well for both linear and non-linear problems. The basic idea of SVM is to find the best hyperplane that separates data points of different classes in the feature space.

In SVM, data points are plotted in a space where each feature is a dimension. The algorithm then identifies the optimal hyperplane that maximizes the margin between different classes. The data points that are closest to the hyperplane and influence its position and orientation are known as support vectors[30].

The mathematical formulation of SVM involves finding the hyperplane that solves the following optimization problem:

$$\min_{w,b} \frac{1}{2}\|w\|^2$$

Subject to:

$$y_i(wx_i + b) \geq 1, \text{ for all } i$$

Here, $w$ is the weight vector, $b$ is the bias, $x_i$ are the training examples, and $y_i$ are the class labels.

For non-linear problems, SVM uses kernel functions to transform the input space into a higher-dimensional space where a linear separation is possible. Common kernels include polynomial, radial basis function (RBF), and sigmoid.

SVM is particularly popular in text classification due to its effectiveness in high-dimensional spaces and its ability to handle overfitting, especially in cases where the number of features exceeds the number of samples[30].

# Data collection

The volume of scientific literature available can make it difficult to identify relevant articles related to artificial intelligence, since researchers typically do not have full access to the entire WoS database. Therefore, we need to propose a search method to retrieve a portion of the WoS database for further analysis. To minimize the number of AI-related articles excluded from our study, it is necessary to ensure our search strategy includes as many AI-related articles as possible.

To begin with, we used the search term 'TS="Artificial Intelligence"' to retrieve all the articles related to AI from the Web of Science database. The publication time range was set from January 1, 2013 to December 31, 2022. The same time range was used for all of our future searches to ensure that we could analyze the AI-related articles from the past 10 years. A total of 95,835 results from Web of Science Core Collection were returned. We then conducted a high-frequency keyword analysis to identify the most common and relevant keywords related to AI. We downloaded the Full records of all the articles returned above, and extracted "Author Keywords" and "Keywords Plus" fields from the records. Those keywords were ranked according to their total times of appearance. We then manually reviewed the keywords which appeared 200 times or more, and discarded those unrelated to AI. When necessary, we consulted wikipedia and WoS search results to clarify the meaning of a certain keyword. Since the aim of this step is to ensure a high recall rate for further classification, we retained a keyword whenever we were not sure if it would bring us more AI-related articles. As a result, a total of 196 keywords were retained.

To further ensure that we were not leaving any important search terms out, we also reviewed the search strategy used in previous studies. As is mentioned above, Liu[21] proposed a comprehensive search strategy for AI-related articles and the result has been widely used, which consists of a core lexical query, two expanded lexical queries, and the WoS category "Artificial Intelligence". We also reviewed all WoS citation topics and determined 10 topics that belong to Artificial intelligence, including Natural Language Processing, Face Recognition, Defect Detection, Reinforcement Learning, Video Summarization, Action Recognition, Object Tracking, Deep Learning, Artificial Intelligence & Machine Learning and Visual Servoing. Our selected search terms (referred to as Tang's approach) were combined with the search terms proposed in Liu's paper. The final search strategy is shown in Table 1. This combined approach allowed us to retrieve a large number of AI-related articles that were relevant to our research objectives.

To make sure our results are comparable with previous results, we used our search strategy in the WoS Science Citation Index Expanded (SCI-Expanded) and Social Sciences Citation Index (SSCI) databases, and set the same time limit January 1, 2013 to December 31, 2022. 662,844 articles were retrieved with Liu's approach. When combined with Tang's approach, there are a total of 2,490,817 articles for further screening and analysis. The results indicate that most of the AI-related articles from WoS cannot be captured by a simple search of TS="Artificial Intelligence".

Table 1. Preliminary search approach for artificial intelligence.

| **Author** | **Search Strategy** | **Search terms** |
|---|---|---|
| Liu et al. | Core lexical query | TS=("Artificial Intelligen*" or "Neural Net*" or "Machine* Learning" or "Expert System$" or "Natural Language Processing" or "Deep Learning" or "Reinforcement Learning" or "Learning Algorithm$" or "Supervised Learning" or "Intelligent Agent") |
| | Expanded lexical query 1 | TS=(("Backpropagation Learning" or "Back-propagation Learning" or "Bp Learning") or ("Backpropagation Algorithm*" or "Back-propagation Algorithm*") or "Long Short-term Memory" or ((Pcnn$ not Pcnnt) or "Pulse Coupled Neural Net*") or "Perceptron$" or ("Neuro-evolution" or Neuroevolution) or "Liquid State Machine*" or "Deep Belief Net*" or ("Radial Basis Function Net*" or Rbfnn* or "Rbf Net*") or "Deep Net*" or Autoencoder* or "Committee |
| | | Machine*" or "Training Algorithm$" or ("Backpropagation Net*" or "Back-propagation Net*" or "Bp Network*") or "Q learning" or "Convolution* Net*" or "Actor-critic Algorithm$" or ("Feedforward Net*" or "Feed-Forward Net*") or "Hopfeld Net*" or Neocognitron* or Xgboost* or "Boltzmann Machine*" or "Activation Function$" or ("Neurodynamic Programming" or "Neuro dynamic Programming") or "Learning Model*" or (Neuro computing or "Neuro-Computing") or "Temporal Difference Learning" or "Echo State* Net*") |
| | Expanded lexical query 2 | TS=("Transfer Learning" or "Gradient Boosting" or "Adversarial Learning" or "Feature Learning" or "Generative Adversarial Net*" or "Representation Learning" or ("Multiagent Learning" or "Multi-agent Learning") or "Reservoir Computing" or "Co-training" or ("Pac Learning" or "Probabl* Approximate* Correct Learning") or "Extreme Learning Machine*" or "Ensemble Learning" or "Machine* Intelligen*" or ("Neuro fuzzy" or "Neurofuzzy") or "Lazy Learning" or ("Multi* instance Learning" or "Multi-instance Learning") or ("Multi* task Learning" or "Multitask Learning") or "Computation* Intelligen*" or "Neural Model*" or ("Multi* label Learning" or "Multilabel Learning") or "Similarity Learning" or "Statistical Relation* Learning" or "Support* Vector* Regression" or "Manifold Regularization" or "Decision Forest*" or "Generalization Error*" or "Transductive Learning" or (Neurorobotic* or "Neuro-robotic*") or "Inductive Logic Programming" or "Natural Language Understanding" or (Ada-boost* or "Adaptive Boosting") or "Incremental Learning" or "Random Forest*" or "Metric Learning" or "Neural Gas" or "Grammatical Inference" or "Support* Vector* Machine*" or ("Multi* label Classification" or "Multilabel Classification") or "Conditional Random Field*" or ("Multi* class Classification" or "Multiclass Classification") or "Mixture Of Expert*" or "Concept* Drift" or "Genetic Programming" or "String Kernel*" or ("Learning To Rank*" or "Machine-learned Ranking") or "Boosting Algorithm$" or "Robot* Learning" or "Relevance Vector* Machine*" or Connectionis* or ("Multi* Kernel$ Learning" or "Multikernel$ Learning") or "Graph Learning" or "Naive bayes* Classif*" or "Rule-based System$" or "Classification Algorithm*" or "Graph* Kernel*" or "Rule* induction" or "Manifold Learning" or "Label Propagation" or "Hypergraph* Learning" or "One class Classif*" or "Intelligent Algorithm*") |
| | WoS category | WC=("Artificial Intelligence") |
| Tang et al. | Lexical query | TS=("action recognition" OR "activation function$" OR "activity recognition" OR "adaboost" OR "AI" OR "algorithm$" OR "anfis" OR "ann" OR "anomaly detection" OR "ant colony optimization" OR "artificial bee colony" OR "artificial neural-network$" OR "artificial-intelligence" OR "attribute reduction" OR "augmented reality" OR "autoencoder$" OR "automa* detection" OR "automa* segmentation" OR "automa* classification" OR "background subtraction" OR "backpropagation" OR "bankruptcy prediction" OR "bayesian network$" OR "bayesian-inference" OR "bidirectional lstm" OR "big data" OR "bootstrap" OR ("brain-computer interface$" OR "brain-computer-interface$") OR "canonical correlation-analysis" OR "cellular neural-network$" OR "classifier$" OR ("cluster-analysis" OR "cluster analysis") OR "cnn" OR "community detection" OR "complex dynamical network$" OR "component analysis" OR "computational intelligen*" OR "computer vision" OR "computer-aided detection" OR "concept drift" OR "consensus model" OR ("convolutional network$" OR "convolutional neural-network$") OR "corpus" OR "crack detection" OR "cross-validation" OR "damage detection" OR "data augmentation" OR "data fusion" OR "data mining" OR "decision tree$" OR "deconvolution" OR "deep neural-network$" OR "defect |
| | | detection" OR "dempster-shafer theory" OR "differential evolution" OR "dimensionality reduction" OR "discriminant-analysis" OR "dynamical network$" OR "edge-detection" OR "eigenface$" OR "emotion recognition" OR "energy minimization" OR "event detection" OR "evidential belief function" OR "evidential reasoning approach" OR "expert-system" OR "exponential stability" OR "exponential synchronization" OR "expression recognition" OR "extended kalman filter" OR "extreme learning-machine$" OR ("face recognition" OR "face-recognition") OR "facial expression recognition" OR "fault-diagnosis" OR "feature subset-selection" OR "feature-extraction" OR "feature-selection" OR "feedforward networks" OR "fuzzy c-means" OR "fuzzy inference system" OR "fuzzy-logic" OR "fuzzy-set$" OR "fuzzy-system$" OR "gan" OR "gaussian process regression" OR "generative adversarial network$" OR "gesture recognition" OR "global exponential stability" OR "gradient descent" OR "grey wolf optimizer" OR "group decision-making" OR "hidden markov-models" OR "human activity recognition" OR "image classification" OR "image registration" OR "image segmentation" OR "image-analysis" OR "imbalanced data" OR "inference system" OR "information extraction" OR "in-silico prediction" OR "intrusion detection" OR "kalman filter" OR "kernel" OR "k-svd" OR "lasso" OR "lda" OR "leader-following consensus" OR "learning framework" OR "learning-based optimization" OR "learning-model" OR "least-squares" OR "linear discriminant-analysis" OR "local binary patterns" OR "logistic-regression" OR "lstm" OR "machine vision" OR "markovian jump systems" OR "metaheuristics" OR "multiagent system$" OR "multilayer feedforward network$" OR "multilayer perceptron" OR "multiobjective optimization" OR "naive bayes" OR "nearest-neighbor" OR "neural-control" OR "neural-network$" OR "nonlinear dimensionality reduction" OR "nonlinear-systems" OR "novelty detection" OR ("object detection" OR "object recognition") OR "object tracking" OR "outlier detection" OR "parameter-estimation" OR "parameter-identification" OR "partial least-squares" OR "particle swarm" OR "pattern-classification" OR "pattern-recognition" OR "pca" OR "pcnn" OR "pedestrian detection" OR "perceptron" OR "permutation entropy" OR "person reidentification" OR "pls" OR "pose estimation" OR "principal component analysis" OR "pso" OR "quantile regression" OR "random forest$" OR "recommender system$" OR "recurrent neural-network$" OR "regression-analysis" OR "regression-models" OR "representation model" OR "robot" OR "robot manipulator$" OR "roc curve" OR "rough set$" OR "rule extraction" OR "scene classification" OR "seizure detection" OR "self-organizing map$" OR "semantic segmentation" OR "semantic similarity" OR "semantic web" OR "sentiment analysis" OR "sequence-based predictor" OR "short-term-memory" OR "smote" OR "sparse representation" OR "species distribution model$" OR ("support vector machine$" OR "support-vector-machine") OR "support vector regression" OR "svm" OR "svr" OR "target detection" OR "text classification" OR "texture analysis" OR "texture classification" OR "time-series" OR "time-varying delay$" OR "traffic flow prediction" OR "trajectory tracking" OR "travel-time prediction" OR "variable selection" OR "variational mode decomposition" OR "visual tracking" OR "visual-attention") |
| | Citation topics | 4.48.672 Natural Language Processing |
| | | 4.17.118 Face Recognition |
| | | 4.17.1950 Defect Detection |
| | | 4.116.862 Reinforcement Learning |
| | | 4.17.1802 Video Summarization |
| | | 4.17.630 Action Recognition |
| | | 4.17.953 Object Tracking |
| | | 4.17.128 Deep Learning |
| | | 4.61 Artificial Intelligence & Machine Learning |
| | | 4.116.2066 Visual Servoing |

The process of data collection is shown in Fig 1.

Fig. 1. Overview of the data collection process

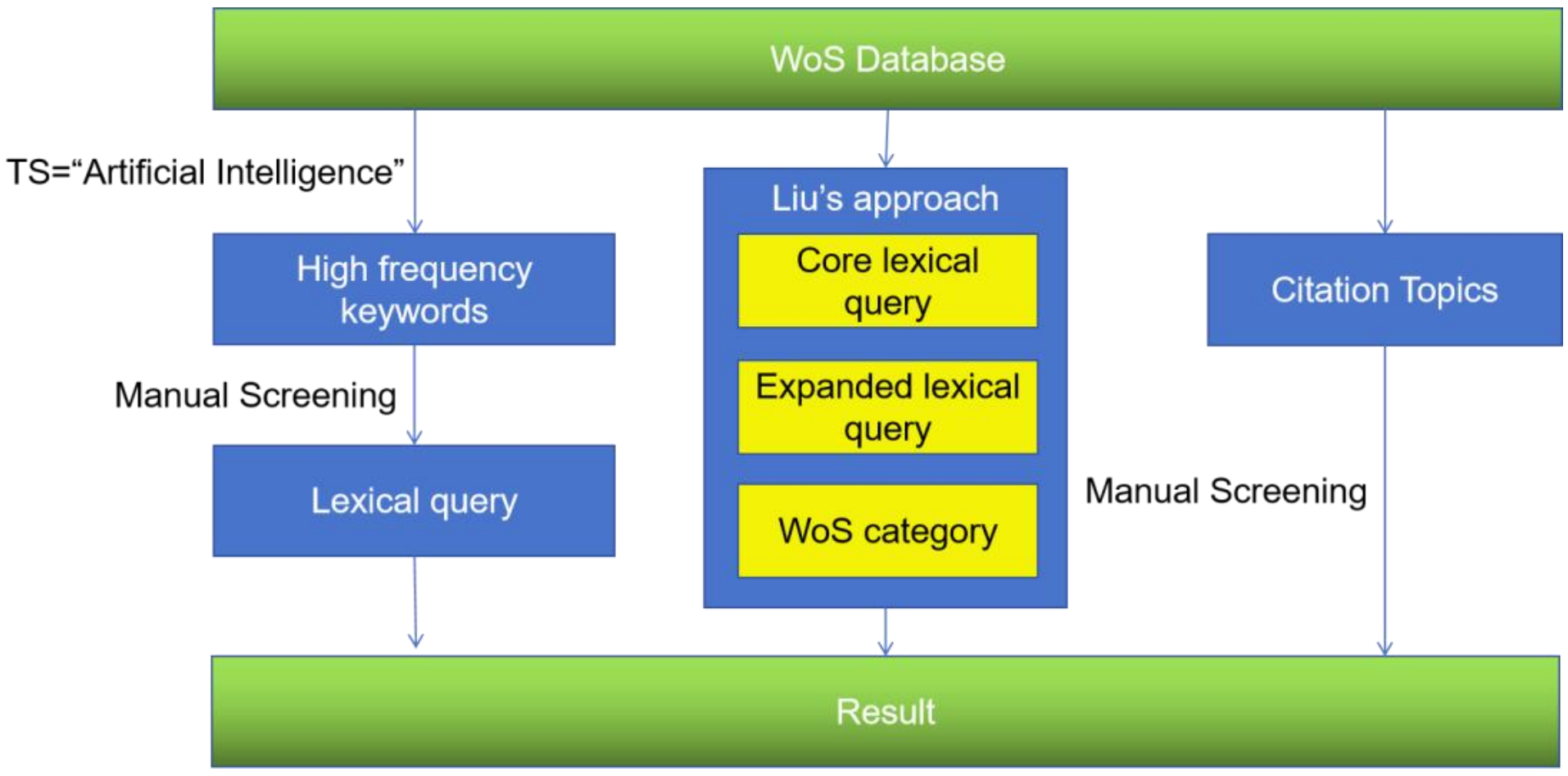

We then downloaded the corpus (referred to as the initial corpus) via the search strategy described in Table 1. In order to obtain the training data, we sampled 4,000 articles from the initial corpus and manually labelled them as "AI-related" and "Other". The definition of AI used in this study is a technology in which some form of statistical learning is employed, and perceives its environment to predict an action that maximizes its chance of achieving its goals[31]. This definition was selected because the article provides very detailed explanation and large number of examples of AI-related technology in its full text. We enrolled two computer science students as coders to independently label the articles. The coders were instructed to follow the aforementioned definition and read through the title and abstract to determine whether an articles is AI-related. The authors reviewed their classification results and discussed with the coders to align their understandings of AI and reduce their differences. Finally, when disagreements between the coders cannot be resolved, an AI expert from Zhejiang university was consulted to determine the final result. We iteratively sampled articles from the initial corpus to keep a balanced class distribution between AI and non-AI. In the end, the positive class was slightly oversampled, resulting in a manually labelled set of 4000 articles in which 50% are positive.

# Model Development

### SciBERT

In our approach to classifying scientific articles, we leveraged SciBERT, a state-of-the-art pretrained language model specifically designed for scientific text. For each article in our manually labeled dataset, which served as the ground truth, we concatenated and lemmatized its title and abstract to form a single input text sequence. We then fine-tuned sciBERT using this concatenated text. 5-fold cross-validation was used in the fine-tuning process. In each fold, four sets were used for training and the remaining set for validation. The output of the SciBERT fine-tuning for our task is a probability score between 0 and 1 for each article, representing the likelihood that the article is AI-related. This is achieved by selecting the sigmoid activation function as the final layer of the sciBERT model.

Let $T$ and $A$ be the title and abstract of an article respectively. The probability score $P$ indicating whether the article is AI-related is obtained by:

$$P = \sigma(W \cdot (SciBERT(Lemmatize(T + A))) + b)$$

$$\sigma(x) = \frac{1}{1+e^{-x}}$$

Here, $\sigma$ is the sigmoid function, $W$ is the weight matrix, and $\mathrm{b}$ is the bias term.

The fine-tuned SciBERT model constitutes the foundation of our article classification system. Building upon this base, we incorporated additional components aimed at enhancing the model's predictive accuracy. These components leveraged other article attributes and were integrated through a ensemble method.

### Decision Tree

In parallel to the fine-tuning of SciBERT, we sought to exploit the categorical metadata associated with scientific articles. The Web of Science (WoS) categories assigned to each article offer a high-level view of the content domain. To make use of this categorical information, we employed a Decision Tree classifier.

This classifier was trained to predict the likelihood of an article being related to artificial intelligence based on its WoS categories. Decision Trees are particularly advantageous for their interpretability and ease of use. To validate the effectiveness of this approach, the 4000 labeled articles from our dataset were subjected to a 5-fold cross-validation scheme, akin to the method applied during the SciBERT fine-tuning process.

At each node $\mathrm{n}$ of the tree, the algorithm selects the best feature $\mathrm{f}$ and a threshold $\theta$ that maximizes Information Gain. This is represented as:

$$Split(n) = \mathrm{argmax}_{f,\theta}\; InformationGain(D, f, \theta)$$

Where Information Gain is calculated as the difference in entropy (measure of uncertainty or randomness) before and after the split. Once the tree is built, the probability of an article being AI-related is estimated based on the proportion of AI-related articles in the leaf node where the article falls.

### Keyword matching

To further refine our classification model, we conducted an in-depth analysis of the high-frequency keywords within the initial corpus. This involved a manual review of the high-frequency keywords in the initial corpus to identify AI-related ones. Our goal was to establish a list of AI-related keywords that could serve as strong indicators for classifying articles as relevant to the field of AI.

For each keyword identified in this process, we randomly selected 20 articles containing the keyword from the initial corpus. This sample was scrutinized not only by the authors but also by external AI experts to ensure that the presence of the keyword was indeed indicative of the article's relevance to artificial intelligence. This validation process confirmed the pertinence of the keyword to the domain of AI and verified that the articles containing these keywords were correctly classified as AI-related. By leveraging expert judgment alongside keyword frequency, we were able to curate a list of keywords with a high degree of confidence in their relevance to artificial intelligence research. The keywords were summarized into the regular expression in Table 2. We then performed regular expression matching on title, abstract and keywords of articles to determine whether it contains any AI-related terms, and the matching results were included into our ensemble model.

Table 2. AI-related keywords used

| **Regular expression** |
|---|
| (Artificial Intelligen.*\|Machine Learning\|Deep Learning\|Genetic Algorithm\|Support Vector Machine\|Image Segmentation\|Particle Swarm\|Reinforcement Learning\|Random Forest\|Computer Vision\|Transfer Learning\|Natural Language Processing\|Supervised Learning\|Semantic Segmentation\|Generative Adversarial Network\|Sentiment Analysis\|Multi-Agent System\|MultiAgent System\|Ensemble Learning\|Extreme Learning\|Recommender System\|Image Retrieval\|Decision Tree\|image Fusion\|Long Short-Term Memory\|Evolutionary Algorithm\|Ant Colony Optimization\|Convolutional Neural Network\|Artificial Neural Network\|Deep Neural Network\|Recurrent Neural Network\|BP Neural Network\|Graph Neural Network\|\bSVM\b\|\bNLP\b\|\bLSTM\b\|\bCNN\b\|\bDNN\b\|\bRNN\b\|\bGNN\b) |

### SVM

Building upon the individual strengths of SciBERT, the decision tree, and keyword matching, we integrate these components using a Support Vector Machine (SVM). The SVM acts as a meta-classifier, taking as input the probability outputs from SciBERT and decision tree, and the binary results from the regular expression keyword matching. The SVM is trained to find an optimal boundary that separates AI-related articles from others. For each article, the input features include the probability score from SciBERT, indicating the probability of the article being AI-related; the output of the decision tree, which predicts the probability based on the article's categorical data; and a binary feature signifying the presence or absence of AI-related terms as determined by our regular expression analysis.

Let $T_x$ , $A_x$ , $K_x$ and $C_x$ represent the title, abstract, and keywords of article $x$ respectively, the SVM function incorporating these specific inputs would be represented as:

$$y = SVM(P_{sciBERT}(T_x + A_x), P_{DT}(C_x), I_{regex}(T_x + A_x + K_x))$$

And $I_{regex}$ is the binary output from the regular expression matching.

Fig. 2. Overview of the model architecture

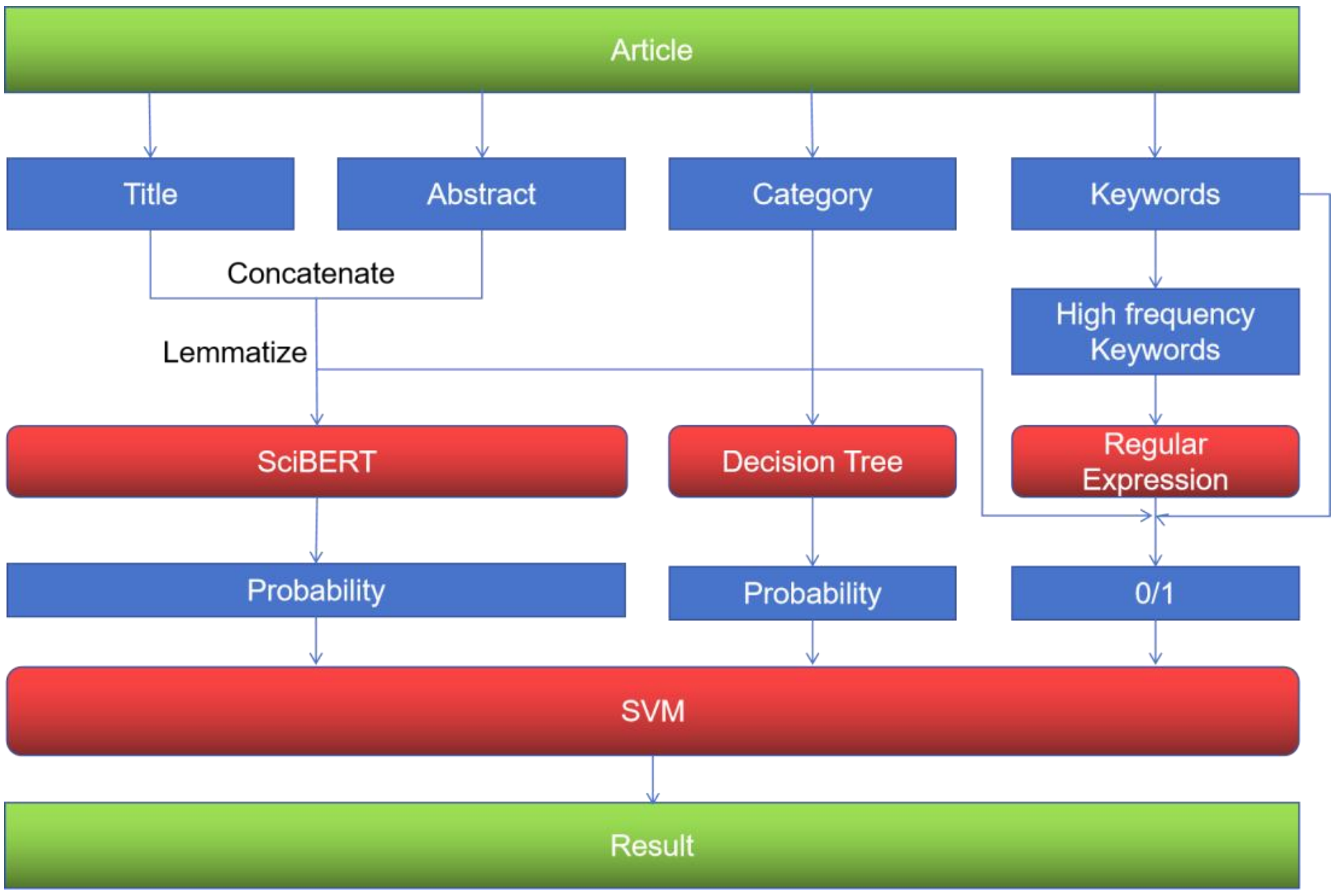

# Results

## Performance evaluation

At the beginning of this section, we examine the overlap between the different approaches used to identify AI-related articles. As is mentioned above, approaches in retrieving AI-related articles range from simple search term "Artificial Intelligence", WoS category "Computer Science, Artificial Intelligence", complicated search terms, to employing deep learning models. Three strategies were selected for further comparison: keyword based search strategy by Liu et al[21] (referred to as Liu's approach), WoS Category="Artificial Intelligence" Approach (referred to as WoS Category approach), and our Ensemble model (referred to as the EnsembleAI model). We selected Liu's search strategy because it achieves the highest performance among existing search strategies[21]. The simplest approach of using a single keyword "Artificial Intelligence" as the query yielded a very low recall, therefore it is omitted in the following analysis. The Venn diagram below illustrates the coverage and intersection of the three approaches.

Fig 3. Comparison of different artificial intelligence classification approaches

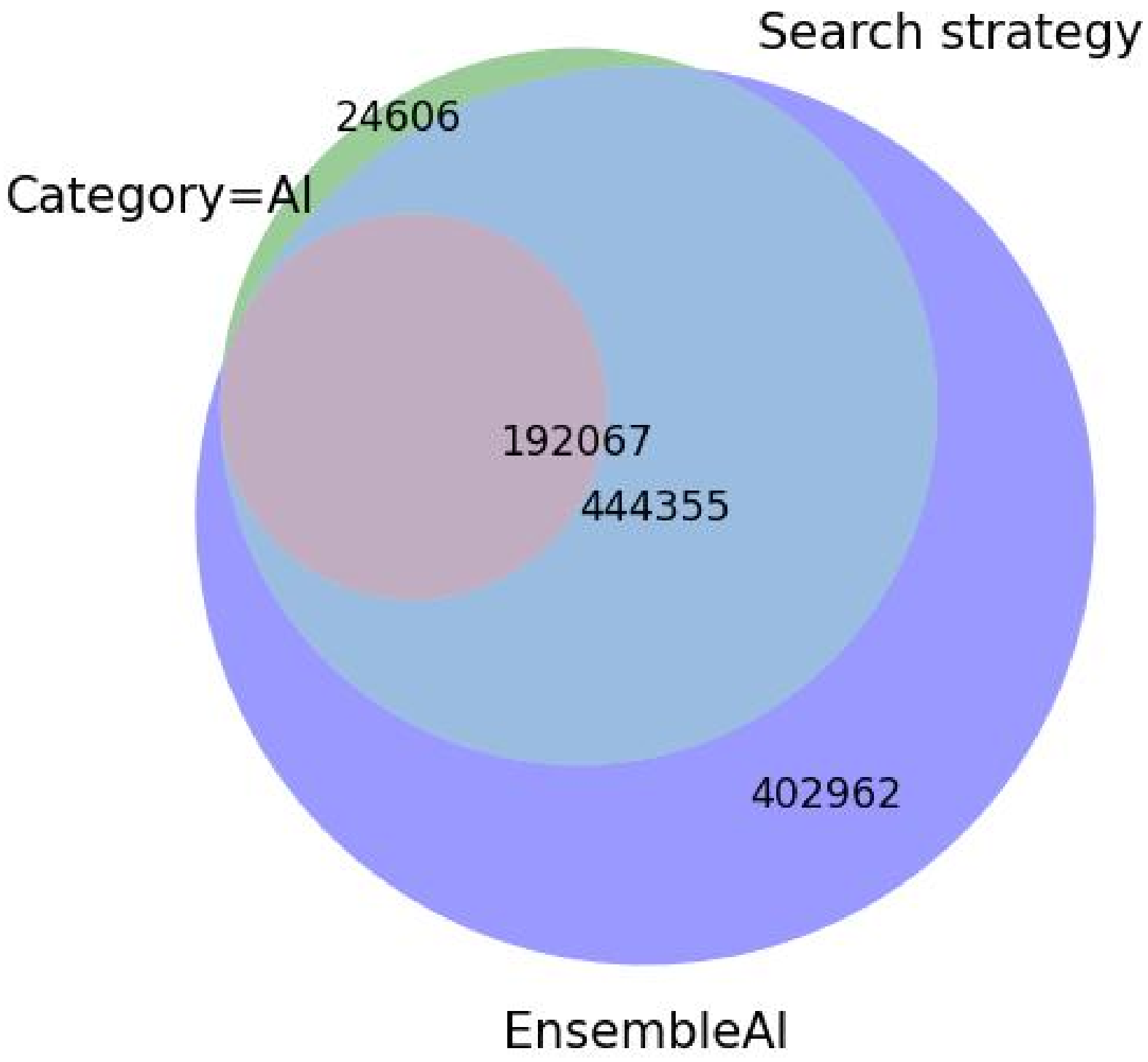

The WoS category approach, which contains 192,067 articles, is included in both Liu's approach and EnsembleAI. Liu's approach is almost included in EnsembleAI, with the exception of 24,606 articles. Our EnsembleAI approach contains the biggest number of articles among the three approaches, with 402,962 unique articles. The significant overlap between the approaches indicate that we share a common understanding of the AI research landscape. However, the disparities in the number of articles identified by Liu's Approach and EnsembleAI suggest that there may be unique aspects captured by our methods.

Next, we evaluated the performance of various methods to identify AI-related articles. Previously manually annotated dataset of 4,000 articles were directly used to evaluate the performance of WoS category and Liu's approach. For EnsembleAI, 5-fold cross validation was employed during training and performance evaluation was done by applying the corresponding test set in each fold and computing the average performance.

The performance of each method is evaluated in terms of precision and recall. Precision measures the fraction of true AI-related articles among those identified by the method, while recall measures the fraction of true AI-related articles identified by the method out of all the AI-related articles in the sample.

The following table summarizes the results:

Table 3. Comparison of the performance of different artificial intelligence classification approaches

| Name | Precision | Recall | F1-score |
|---|---|---|---|
| WoS category | 0.934 | 0.325 | 0.482 |
| Liu’s approach | 0.980 | 0.669 | 0.795 |
| EnsembleAI | 0.915 | 0.973 | 0.943 |

Table 3 presents the performance metrics of three distinct approaches for classifying articles as related to artificial intelligence. Precision measures the proportion of true positive results among all positive cases identified by the classifier, while recall (or sensitivity) measures the proportion of true positive results among all actual positive cases. The F1-score is the harmonic mean of precision and recall, providing a single metric that balances both.

The WoS category-based approach shows precision of 0.934, since not all texts in this category were classified as AI related by our experts. For example, articles dealing with human-machine interfaces, chemometrics and bilevel optimization were deemed non-AI by our experts, and we excluded them from the AI-related articles. Its recall is quite low (0.325), suggesting it fails to identify a large number of AI-related articles. Consequently, its F1-score is relatively low (0.482), reflecting the imbalance between precision and recall.

Liu’s approach significantly improves upon the WoS category, particularly in recall (0.669), while maintaining excellent precision (0.980). This balance results in a much higher F1-score (0.795), suggesting a more effective model overall.

The EnsembleAI shows a slight decrease in precision (0.915) compared to Liu’s approach, but has a superior recall (0.973). This high recall indicates that EnsembleAI is particularly adept at identifying most AI-related articles. Despite the slight trade-off in precision, EnsembleAI achieves the highest F1-score (0.943) among the three methods, indicating a robust performance and a well-balanced approach between precision and recall.

**Ablation study**

The ablation study will methodically examine the EnsembleAI model by selectively removing or isolating specific components, such as the decision tree, sciBERT, keyword matching, and the SVM integrator. This process aims to reveal the impact and value of each component within the ensemble framework. Through this study, we intend to quantify the contribution of each part of the EnsembleAI model to its precision, recall, and F1-score.

Overall, the ablation study highlights the synergistic effect of the combined components in the EnsembleAI model. Each component contributes uniquely to the model's performance, with the decision tree and abstract elements being particularly crucial for maintaining high recall, and the SVM integrator more significantly affecting precision. The balanced performance of the original EnsembleAI model underscores the importance of integrating multiple approaches for effective identification of AI-related articles.

Table 4. Comparison of the performance of models with different fields as input

| | EnsembleAI | EnsembleAI-Decision Tree | EnsembleAI-Keywords | EnsembleAI-SVM | EnsembleAI-title | EnsembleAI-abstract |
|---|---|---|---|---|---|---|
| precision | 0.915 | 0.921 | 0.915 | 0.902 | 0.905 | 0.930 |
| recall | 0.973 | 0.914 | 0.959 | 0.978 | 0.970 | 0.936 |
| F1-score | 0.943 | 0.918 | 0.937 | 0.939 | 0.936 | 0.933 |

The data from the ablation study of the EnsembleAI model provides insightful revelations about the contributions of its individual components to the overall performance in identifying AI-related articles. Here's an analysis of each variant:

EnsembleAI (Original Model): With a precision of 0.915, recall of 0.973, and an F1-score of 0.943, the original EnsembleAI model exhibits a well-balanced performance.

EnsembleAI-Decision Tree: Removing the decision tree component slightly increases precision to 0.921 but notably reduces recall to 0.914. The F1-score drops to 0.918.

EnsembleAI-Keywords: The removal of keyword matching maintains the same precision (0.915) but slightly decreases recall to 0.959. The F1-score also sees a slight decrease to 0.937.

EnsembleAI-SVM: Excluding the SVM integrator decreases precision to 0.902 and slightly increases recall to 0.978. The F1-score is relatively stable at 0.939. Removing SVM is done by retaining the results from sciBERT and decision tree with a probability of 0.5 and more, and computing the union of positive results from sciBERT, decision tree and keywords matching. This method outperformed majority voting and soft voting, therefore it was selected as the way to replace the SVM integrator.

EnsembleAI-title: Removing the title from the SciBERT input lowers precision slightly to 0.905 and recall to 0.970, with an F1-score of 0.936.

EnsembleAI-abstract: Excluding the abstract from SciBERT input results in the highest precision (0.930) but lowers recall to 0.936, leading to an F1-score of 0.933.

Furthermore, we explored the effectiveness of using only the SciBERT component, as well as the impact of replacing SciBERT with other BERT variants when classifying AI-related articles.

BERT-base, developed by Google, was one of the first models to use Transformer architecture, significantly changing the landscape of text-based analysis by providing a deep, bidirectional understanding of context within language. It is pre-trained on a vast corpus of text and designed to be fine-tuned for a variety of NLP tasks[26].

RoBERTa, which stands for Robustly Optimized BERT Pretraining Approach, is a model developed by Facebook AI that builds upon BERT's architecture. It has been optimized with more data, larger batch sizes, and longer training times, leading to improved performance over BERT-base in many benchmarks[32].

BERT-base and RoBERTa were selected as alternative models and we conducted a series of additional experiments of feeding different combinations of textual data into these models, including title + abstract, and concatenated fields of title + abstract + WoS categories + keywords. The results are shown in Table 5.

Table 5. Comparison of the performance of models with different fields as input

| | EnsembleAI | sciBERT(title +abstract) | BERT-base (title+abstract) | RoBERTa (title+abstract) | sciBERT(title+abstract +category+keywords) |
|---|---|---|---|---|---|
| precision | 0.915 | 0.928 | 0.919 | 0.911 | 0.935 |
| recall | 0.973 | 0.896 | 0.877 | 0.879 | 0.904 |
| F1-score | 0.943 | 0.911 | 0.898 | 0.895 | 0.919 |

EnsembleAI: This is the original EnsembleAI model with a precision of 0.915, recall of 0.973, and F1-score of 0.943. It serves as the baseline for comparison with other model configurations.

sciBERT(title+abstract): By only utilizing sciBERT with the titles and abstracts, we see a slight increase in precision to 0.928 but a notable drop in recall to 0.896. The F1-score of 0.911 indicates that while sciBERT alone is effective, the additional components in the EnsembleAI model contribute to overall performance.

BERT-base (title+abstract): Replacing sciBERT with BERT-base leads to a precision of 0.919 and a further decrease in recall to 0.877. The F1-score drops to 0.898, suggesting that BERT-base is less effective at identifying AI-related articles than the tailored sciBERT in the context of this task.

RoBERTa (title+abstract): With RoBERTa, there is a slight decrease in both precision and recall compared to BERT-base, resulting in an F1-score of 0.895.

sciBERT(title+abstract+category+keywords): This configuration, which feeds a concatenation of title, abstract, WoS category, and keywords directly into sciBERT, achieves a precision of 0.935 and a recall of 0.904. The F1-score improves to 0.919, indicating that incorporating category and keywords information directly into sciBERT's input can enhance performance, though it does not reach the efficacy of the full EnsembleAI model.

These results affirm that while sciBERT is a powerful tool for classifying AI-related articles, the combination of decision trees, keyword matching, and an SVM integrator in the EnsembleAI model leads to superior recall and the best overall performance as measured by the F1-score. Each component of the EnsembleAI model contributes uniquely to its high level of accuracy and ability to comprehensively identify relevant articles, illustrating the benefits of a multifaceted approach to article classification in rapidly evolving research fields like AI.

**Publication Volume Trends**

The analysis of publication trends for artificial intelligence (AI) articles demonstrates a continuous growth pattern from 2013 to 2022. This growth follows an exponential trajectory, with a marked acceleration from 2015 onwards, reaching a peak in 2022. This surge in AI research publications is evident in the increase in both the number of AI-related publications and their percentage in the Web of Science (WoS) database. As of 2022, we estimate that more than 7% publications in that year are associated with AI. Meanwhile, the total number of articles in the WoS category "Artificial Intelligence" published each year increased moderately, and slightly declined in the last year. This indicates a rapid growth of interdisciplinarity in the field of artificial intelligence.

Fig 4. AI-related publications and percentage of publications by year

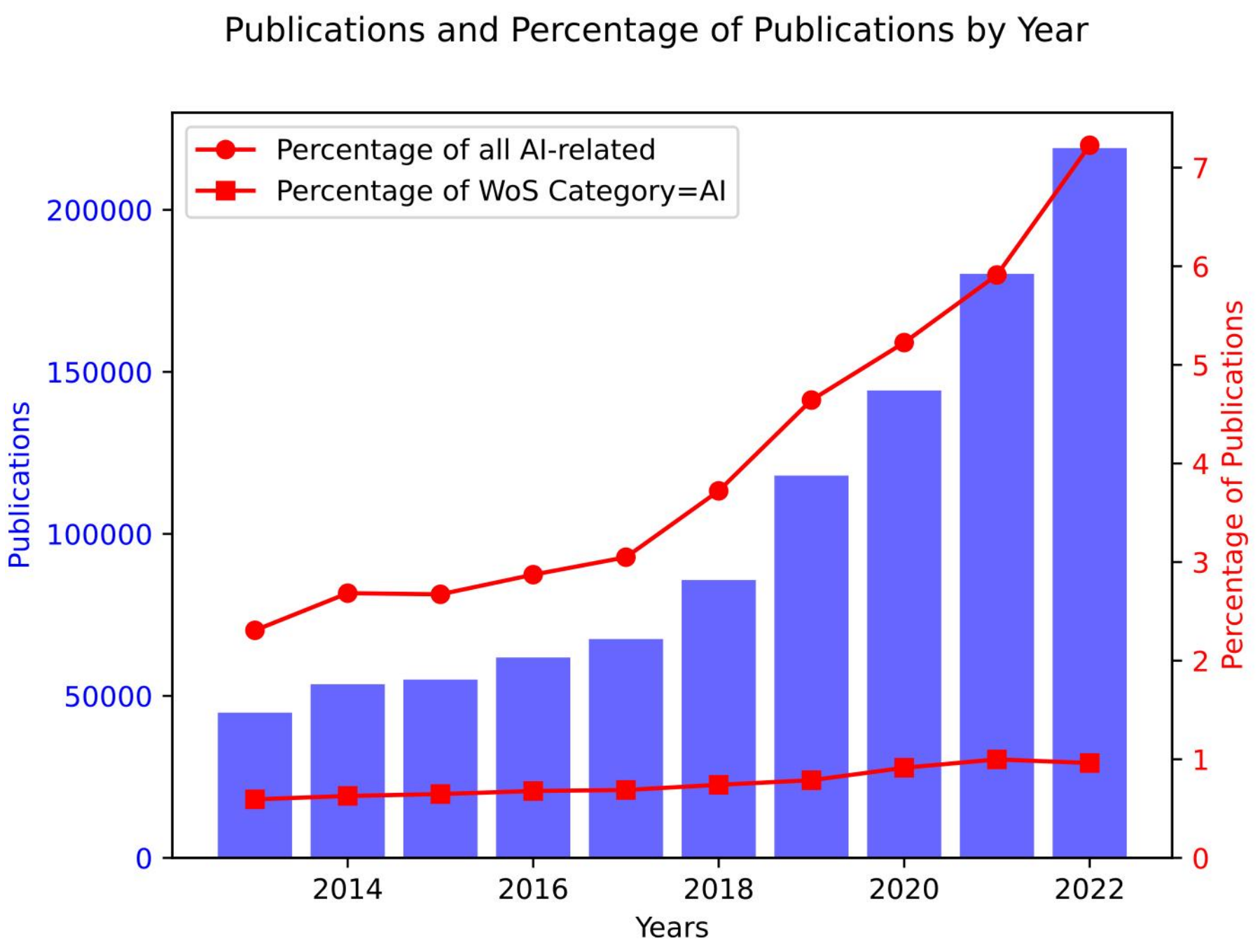

Researchers from around the world contribute to the AI publication landscape, with the dataset including articles from numerous countries. However, the majority of publication output is associated with a small group of leading countries. The top ten countries contribute to more than 70% of the total worldwide AI articles published in the period 2013-2022. China and the US emerge as the most productive countries in terms of AI publications, followed by India, South Korea, England, Germany, Iran, Spain, Japan, and Canada.

The distribution of AI publications by year reveals a consistent increase in the number of AI-related publications. This growth is particularly notable for China, witnessing a rapid increase in the number and world share of AI publications from 2013 to 2022. While the US has maintained a steady pace of growth in terms of total publication count, its world share has been slowly dropping. India ranks third globally, with a steadily increasing number and world share during the past decade. South Korea's growth has been impressive as well, with the number of publications almost sextupling between 2013 and 2022. European countries like England, Germany, Spain, and Italy have experienced more modest growth in scientific publications. Although these countries continue to contribute significantly to global scientific output, their growth rates are relatively lower compared to countries like China and India.

Fig 5. Annual AI-related publications by country

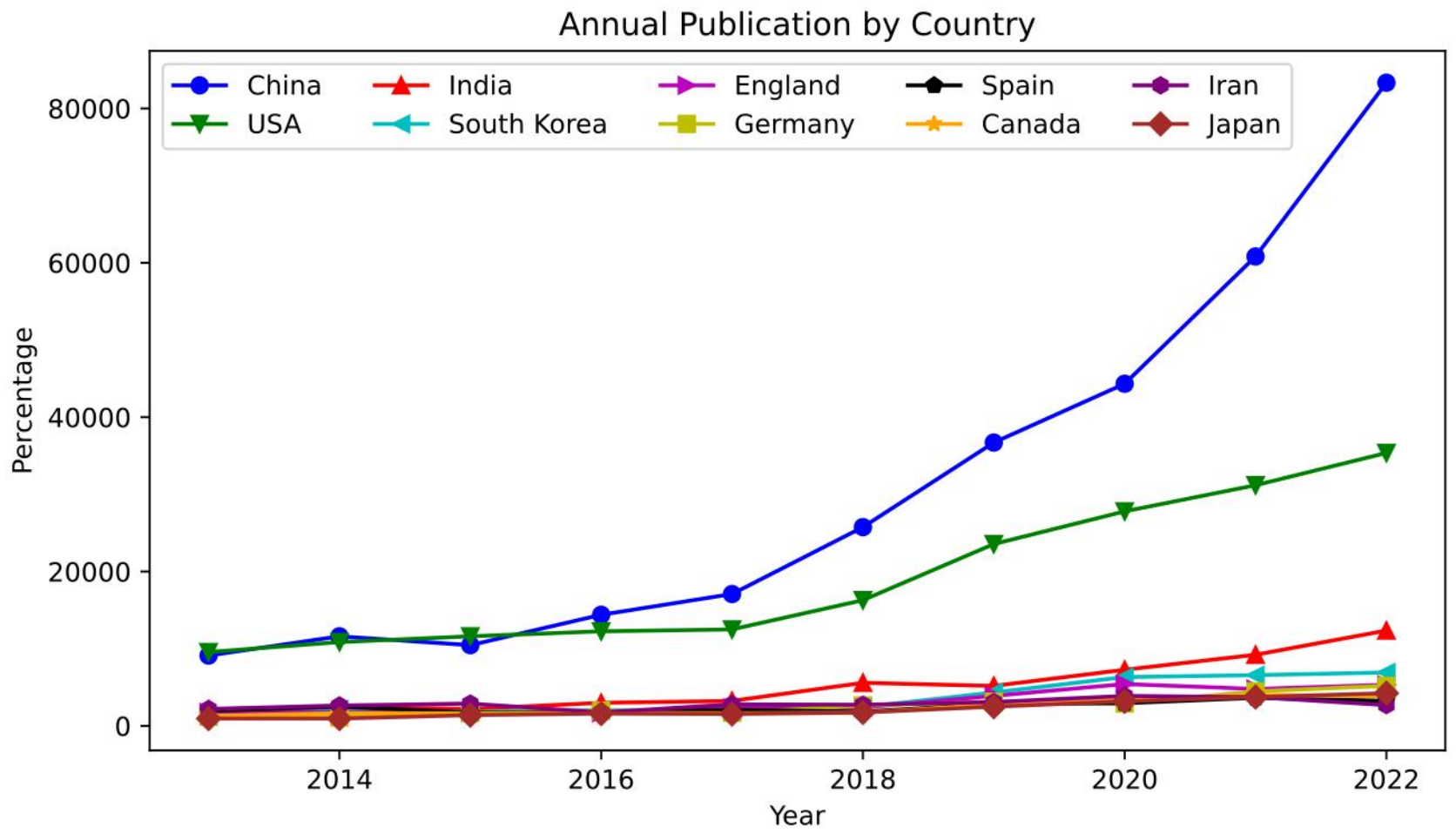

Fig 6. Annual world share of AI-related publications by country

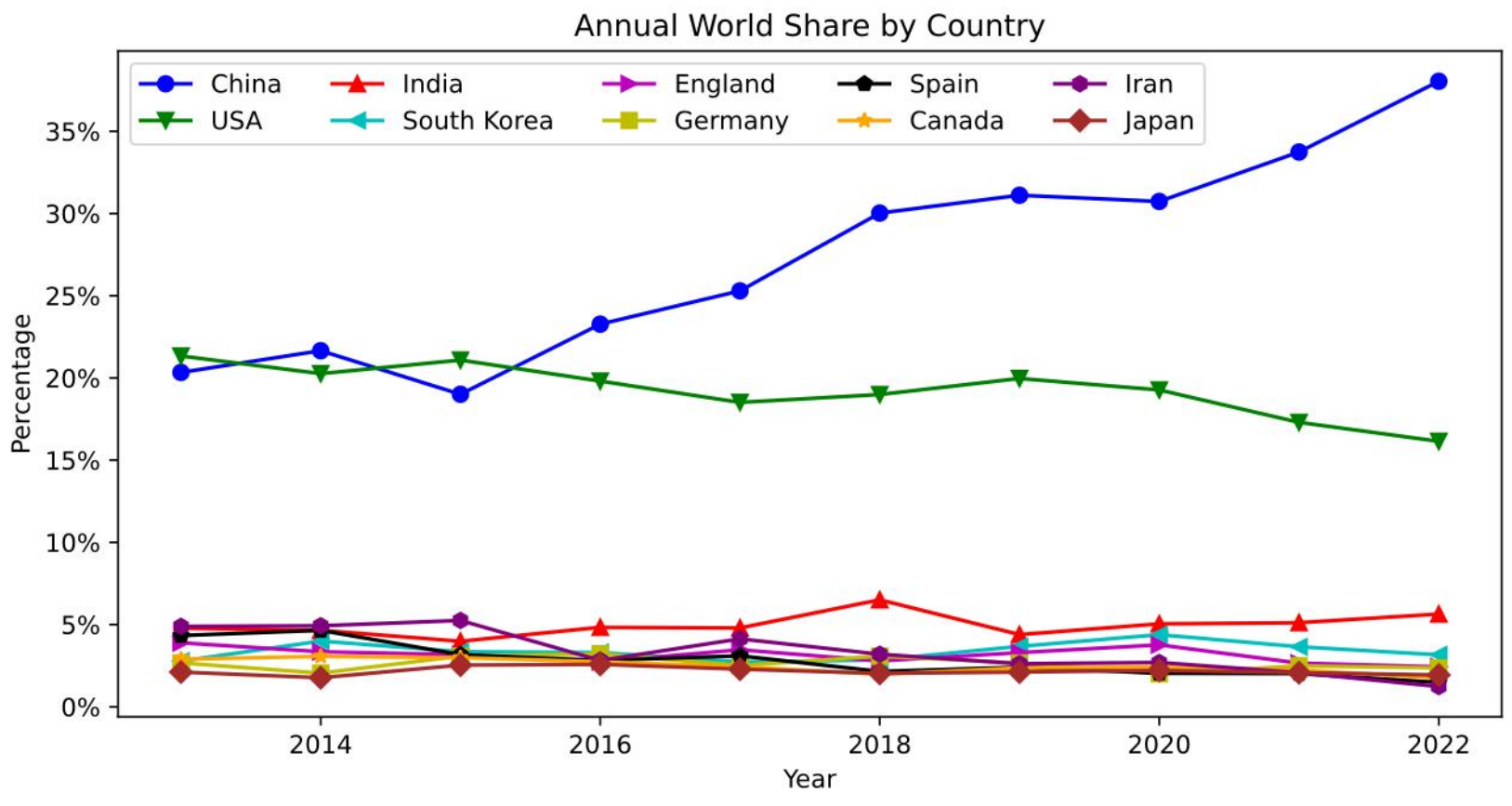

**Major Research Themes**

In this section, we discuss the major research themes in AI by examining the high-frequency keywords and research subfields in the AI-related publications. The analysis of keywords provides insights into the prominent topics and trends in the field. Here are the top 20 high-frequency keywords and their occurrence in the publications:

Table 4. High-frequency keywords in AI-related publications

| Rank | Research field | Occurrence |
|---|---|---|
| 1 | Machine learning | 65,101 |
| 2 | Deep learning | 55,687 |
| 3 | Neural network | 25,403 |
| 4 | Feature extraction | 23,759 |
| 5 | Convolutional neural network | 21,169 |
| 6 | Optimization | 21,165 |
| 7 | Artificial intelligence | 20,023 |
| 8 | Artificial neural network | 19,525 |
| 9 | Genetic algorithm | 14,644 |
| 10 | Classification | 13,797 |
| 11 | Training | 13,000 |
| 12 | Task analysis | 12,701 |
| 13 | Image segmentation | 8,218 |
| 14 | Support vector machine | 7,969 |
| 15 | Feature selection | 7,820 |
| 16 | Particle swarm optimization | 7,675 |
| 17 | Reinforcement learning | 7,670 |
| 18 | Fault diagnosis | 7,421 |
| 19 | Random forest | 7,073 |
| 20 | Data mining | 6.873 |

These high-frequency keywords reveal the core focus areas in AI research, such as machine learning and deep learning. Techniques such as feature extraction, classification, and optimization are central to AI research. Additionally, the prominence of various neural network architectures, including convolutional neural networks, artificial neural networks, and reinforcement learning, indicates the importance of these models in the field.

The analysis also highlights the use of specific algorithms, such as support vector machines, genetic algorithms, random forests, and feature selection techniques. Data mining, a technique used to analyze and extract patterns from large datasets, also emerges as a significant theme in AI research.

To gain a deeper understanding of the strengths and weaknesses of our ensembleAI approach, we will conduct a comparative analysis between our ensemble model and the search term-based approach, specifically looking at their high-frequency keywords. As is mentioned before, Liu's approach was selected to represent the traditional search-term based strategy because of its relatively high performance. The high frequency keywords of Liu's approach is shown in Table 5.

Table 5. High-frequency keywords in AI-related publications (Liu's approach)

| Rank | Research field | Occurrence |
|---|---|---|
| 1 | Machine learning | 66,645 |
| 2 | Deep learning | 55,936 |
| 3 | Neural network | 26,100 |
| 4 | Convolutional neural network | 21,175 |
| 5 | Artificial intelligence | 20,571 |
| 6 | Artificial neural network | 19,575 |
| 7 | Feature extraction | 18,927 |
| 8 | Classification | 11,804 |
| 9 | Training | 11,406 |
| 10 | Task analysis | 9,762 |
| 11 | Optimization | 8,617 |
| 12 | Support vector machine | 8,019 |
| 13 | Reinforcement learning | 7,770 |
| 14 | Random forest | 7,471 |
| 15 | Feature selection | 6,375 |
| 16 | Transfer learning | 6,276 |
| 17 | Natural language processing | 6,126 |
| 18 | Generative adversarial network | 5,429 |
| 19 | Data mining | 5,329 |
| 20 | Image segmentation | 5,180 |

The ensemble model and the search term-based approach in AI research share several similarities, particularly in their emphasis on core areas such as machine learning, deep learning, and various types of neural networks. Both methodologies underscore the significance of these fields, with "Machine Learning" and "Deep Learning" consistently appearing at the top in both models. This highlights a universal recognition of these areas as fundamental to AI research.

Upon closer examination, the ensemble model tends to have a higher occurrence of certain keywords compared to the search term-based approach. For example, "Feature Extraction" appears more prominently in the ensemble model. This could suggest a greater focus or reliance on this aspect in the methodologies or research papers analyzed by the ensemble model. In addition, keywords related to optimization algorithms, including "Optimization", "Genetic algorithm", and "Particle swarm optimization" ranks higher in the ensemble model. This tendency indicates that the ensemble model is more adept at identifying and highlighting articles focused on optimization algorithms.

Next, we will compare the high-frequency Web of Science (WoS) categories between the two approaches - our ensemble model and the search term-based approach. This comparison aims to highlight the differences in thematic focus within AI research as captured by these two distinct methodologies.

Fig 7. Annual AI-related publications by WoS category

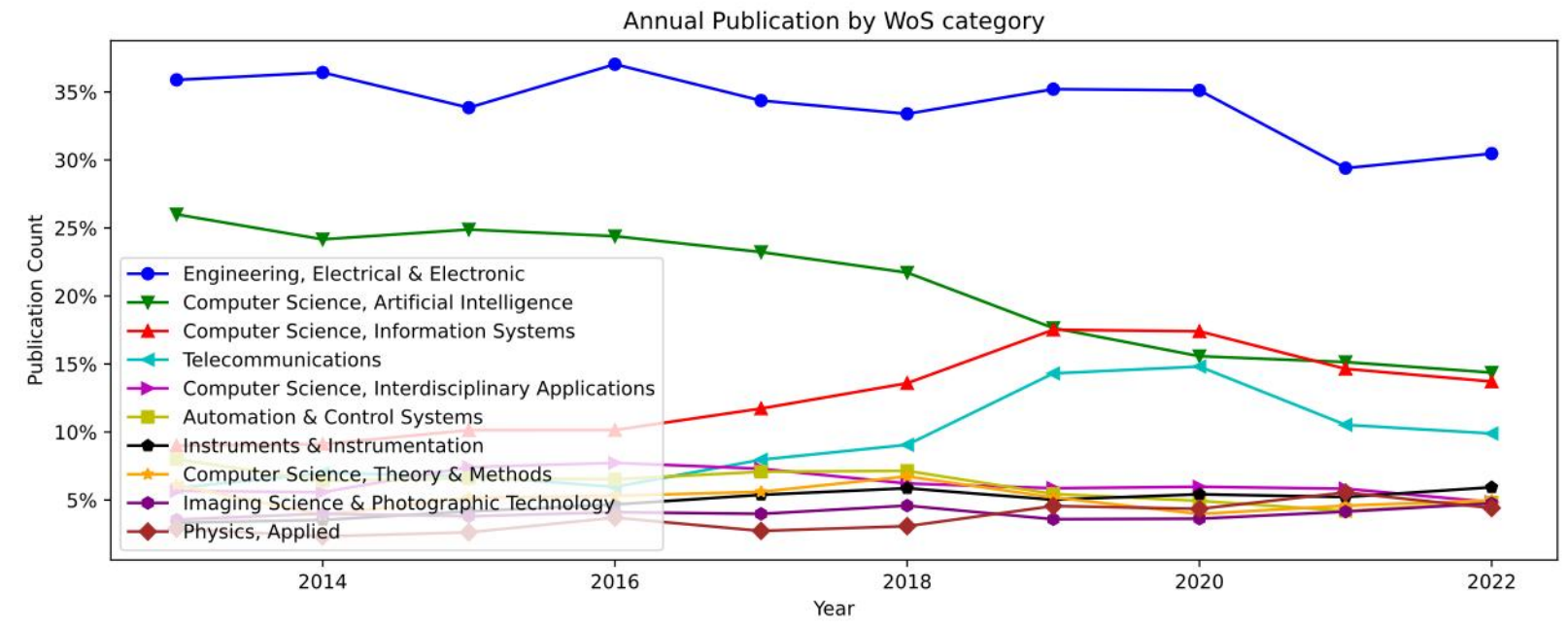

Discipline-wise, AI research has been particularly prominent in fields such as Engineering (Electrical & Electronic), Computer Science (Artificial Intelligence), Information Systems and Telecommunication, in terms of WoS category. We noted that the category of 'Artificial Intelligence' itself ranks only second in the chart, making up only 14.4% of all the AI-related publications in 2022, and its share has been dropping during the last decade. The distribution of AI publications across these disciplines has generally grown over the years, showcasing the interdisciplinary nature of AI research.

Fig 8. Annual AI-related publications by WoS category (Liu's approach)

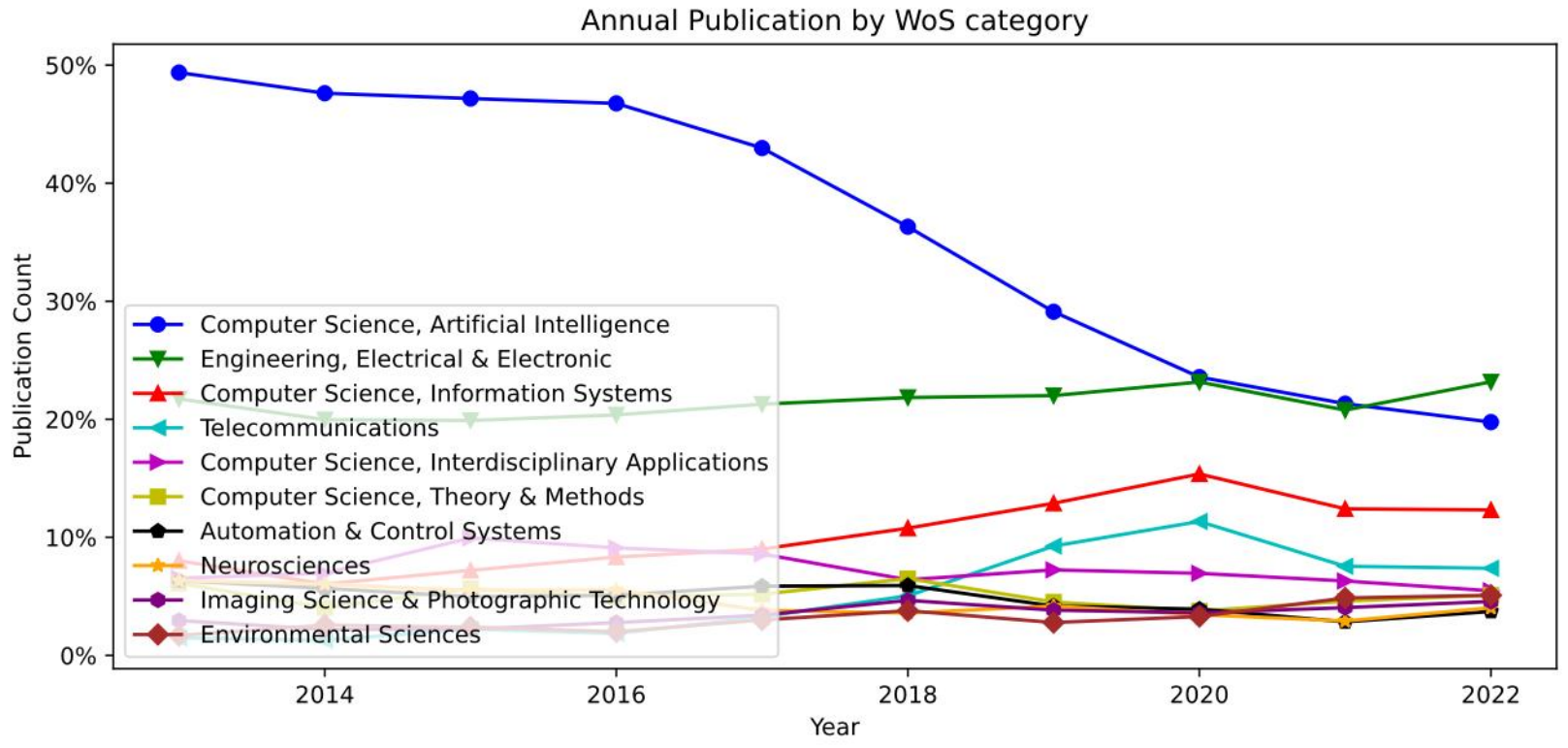

The search term-based approach yielded similar results, with the same top categories: Engineering (Electrical & Electronic), Computer Science (Artificial Intelligence), Information Systems, and Telecommunication. However, in this approach, the Computer Science (Artificial Intelligence) category held a much larger share, nearly 50% in 2013, only falling to the second highest in 2022. Unlike the search term-based approach, which heavily relies on the AI category, the ensemble model appears to provide a broader, more inclusive representation of AI research across various disciplines, effectively showcasing its interdisciplinary reach.

## Discussion

In this study, we proposed a combined approach that integrates a decision tree, SciBERT, and keyword matching across various fields, followed by the use of a Support Vector Machine (SVM) to amalgamate the results. We then identified scholarly articles belonging to the field of artificial intelligence, and analyzed research trends in the field. The performance of our approach was evaluated with manually labeled data from the initial corpus. The results indicate that we achieved significantly increased recall and similar precision, when compared with strategies from previous research.

We noted that previous research has employed machine learning on classifying AI research articles[22, 24]. Siebert et al.[24] achieved a self-reported accuracy of 85%, but when applied to arXiv data, Dunham et al.[22] estimated their approach has a precision of 74% and recall of 49%. Dunham et al.'s approach had a precision of 83% and recall of 85%, but their estimation is based on existing categories. Thus, the fact that only a portion of AI-related publications belong to the 'Artificial Intelligence' category would make the actual performance significantly lower than the reported values. In conclusion, our approach has the best performance among the approaches employing machine learning to classify AI-related research articles.

While our study provides valuable insights, there are limitations to our approach. As is mentioned earlier, the concept of artificial intelligence itself lacks a clear and widely accepted definition[33]. Previous definitions vary from "The science of making machines do things that would require intelligence if done by men"[34] to "the endowment of machines with human-like capabilities through simulating human consciousness and thinking processes using advanced algorithms or models"[35]. In this study, we selected the definition with the most detailed explanation and the highest number of specific examples to our knowledge. We tried using different definitions, and the performance evaluation results can vary. We labeled a small data set with different definitions and the Cohen's Kappa appeared to be around 0.9, indicating a very high level of agreement. Therefore, the differences were small and did not affect our conclusions.

In addition to title, abstract, keywords and WoS categories, we also attempted to include other fields such as journal name or author name into the model. But results showed that they were weak indicators of AI relevance, and including them into the SVM would lower the performance. They were therefore excluded in this study. But with more advanced modelling and feature engineering techniques, they may prove useful in the classification of research articles in future research.

In light of the growing importance of AI, future work can extend our approach to the analysis of patent data in AI and text-mining of websites. This proposition is supported by the findings of previous studies. Researchers have conducted a landscape analysis of AI innovation dynamics and technology evolution using a new AI patent search strategy, incorporating patent analyses, network analyses, and source path link count algorithms[36]. Our ensemble approach can be applied to further enhance the understanding of AI patenting trends and cross-organization knowledge flows. Another study[37] employed topic modeling, a text-mining approach, on archived website data to investigate sales growth for green goods enterprises. This study demonstrated the potential of website data to gauge internal capabilities and market responsiveness. By utilizing the natural language processing ability of the ensemble models to enhance text-mining performance, future work can unlock new insights into the strategic management of innovation and entrepreneurship. Furthermore, our approach can be directly applied to other fields of research, such as nanotechnology[38], synthetic biology[17] and cancer research[39].

## Conclusion

In conclusion, this study presents an ensemble approach to address the challenges of text identification in the rapidly evolving field of Artificial Intelligence. Our approach demonstrates a high precision of 92% and successfully captures about 97% of AI-related articles in the Web of Science (WoS) corpus.

By comparing our approach with existing search-term based methods, we conclude that our approach yielded a similar precision and significantly increased recall, and resulted in a 0.15 increase in F1 score. While the two approaches were similar in common research themes, our ensemble approach revealed higher interdisciplinarity and was able to identify more articles in the topic of feature extraction and optimization.

Our analysis reveals exponential growth in AI research publications, particularly since 2015, with an increasing level of interdisciplinarity. This trend underscores the continued expansion and influence of AI research, highlighting the increasing interest in various subfields and their applications across multiple sectors.

This study highlights the potential of our comprehensive approach in facilitating accurate text identification and analysis in emerging research fields like AI. By incorporating ensemble models into the text-mining process of different data sources, various stakeholders, such as researchers, policymakers, and industry practitioners, can be enabled to leverage the power of deep learning models for future research, policy decisions, and technological advancements.

## Acknowledgments

We thank professor Philip Shapira, Dr. James Dunham and Lizhou Fan for their discussion.